\documentclass[conference]{IEEEtran}
\IEEEoverridecommandlockouts                              

\usepackage{graphicx}          
\usepackage{todonotes}

\usepackage{amsmath}
\usepackage{amssymb}
\usepackage{adjustbox}
\usepackage{aligned-overset}
\usepackage{arydshln}
\usepackage{calc}
\usepackage{cancel}
\usepackage{changepage}		
\usepackage{color}
\usepackage{extarrows}
\usepackage{fancyref}
\usepackage{float}
\usepackage{graphicx}
\usepackage[draft=true,bookmarks=false]{hyperref}
\usepackage{mathtools}
\usepackage{matlab-prettifier}
\usepackage{MnSymbol}
\usepackage{multicol}
\usepackage{newunicodechar}
\usepackage[normalem]{ulem} 
\usepackage{ragged2e}
\usepackage{soul}
\usepackage{units}
\usepackage{subcaption}

\def\BibTeX{{\rm B\kern-.05em{\sc i\kern-.025em b}\kern-.08em
    T\kern-.1667em\lower.7ex\hbox{E}\kern-.125emX}}

\renewcommand{\vec}[1]{\mathbf{#1}}
\newcommand{\gvec}[1]{\boldsymbol{#1}}
\newcommand{\eig}{\operatorname{eig}}
\renewcommand{\(}{\left(}
\renewcommand{\)}{\right)}
\newcommand{\R}{\mathbb{R}}
\lstMakeShortInline[style=Matlab-editor]"

	\usepackage{ifthen}
	\newboolean{IncludeAppendix}   
	\setboolean{IncludeAppendix}{true}
	\newcommand{\IncludeAppendix}[1]{\ifthenelse{\boolean{IncludeAppendix}} {#1}{} }	

	\usepackage{ifthen}
	\newboolean{draft}   
	\setboolean{draft}{false}
	\ifthenelse{\boolean{draft}}{
	}
	{
		\setuptodonotes{disable}
	}
	\newcommand{\form}[1]{\ifthenelse{\boolean{draft}}{\textcolor{orange}{#1}}{#1}}

\title{\LARGE \bf
Robust Model Predictive Longitudinal Position Tracking Control for an Autonomous Vehicle Based on Multiple Models
}

\author{
\IEEEauthorblockN{1\textsuperscript{st} Andr\'{e} Kempf}
\IEEEauthorblockA{\textit{Research \& Development (RD/AFM)}\\ 
\textit{Mercedes-Benz AG}\\
Stuttgart, Germany \\
andre.a.kempf@daimler.com}
\and
\IEEEauthorblockN{2\textsuperscript{nd} Markus Herrmann-Wicklmayr}
\IEEEauthorblockA{\textit{Chair of Systems Modeling and Simulation} \\
\textit{University of Saarland}\\
Saarbruecken, Germany \\
markus.herrmannwicklmayr@uni-saarland.de}
\and
\IEEEauthorblockN{3\textsuperscript{rd} Steffen M\"{u}ller}
\IEEEauthorblockA{\textit{Department of Automotive Engineering} \\
\textit{Technical University of Berlin}\\
Berlin, Germany \\
steffen.mueller@tu-berlin.de}
}

\begin{document}

\IEEEoverridecommandlockouts
\IEEEpubid{\makebox[\columnwidth]
{978-1-7281-2547-3/20/\$31.00~\copyright2020 IEEE \hfill} 
\hspace{\columnsep}\makebox[\columnwidth]{ }}
\maketitle
\IEEEpubidadjcol

\begin{abstract}

The aim of this work is to control the longitudinal position of an autonomous vehicle with an internal combustion engine. The powertrain has an inherent dead-time characteristic and constraints on physical states apply since the vehicle is neither able to accelerate arbitrarily strong, nor to drive arbitrarily fast. A model predictive controller (MPC) is able to cope with both of the aforementioned system properties. MPC heavily relies on a model and therefore a strategy on how to obtain multiple linear state space prediction models of the nonlinear system via input/output data system identification from acceleration data is given. The models are identified in different regions of the vehicle dynamics in order to obtain more accurate predictions. The still remaining plant-model mismatch can be expressed as an additive disturbance which can be handled through robust control theory. Therefore modifications to the models for applying robust MPC tracking control theory are described. Then a controller which guarantees robust constraint satisfaction and recursive feasibility is designed. As a next step, modifications to apply the controller on multiple models are discussed. In this context, a model switching strategy is provided and theoretical and computational limitations are pointed out. Lastly, simulation results are presented and discussed, including computational load when switching between systems.

\end{abstract}

\section{INTRODUCTION}

Fully autonomous driving of passenger vehicles is on the edge of becoming a daily experience in some countries. While there has been an extensive amount of research within the last century on this topic, promising practical implementations start to appear just now. This is especially thanks to advances in algorithms, computing and sensor technology. One field which is important for getting autonomous vehicles on the road is vehicle control \cite{watzenig2017, hima2011}. 

The control problem can be split in different categories, depending on the control architecture. There exist solutions for combined lateral and longitudinal control of the vehicle \cite{falcone2010,hess2013,katriniok2013,menhour2011}, but also approaches for separated architectures can be found in literature \cite{attiaLongitudinalControlAutomated2012,calzolari2017,raffo2009}. In the following, we will focus on independent longitudinal control. One reason for this is that a combination of lateral and longitudinal control becomes crucial at driving limits, which are not within the scope of this work. Another reason is, that an independent concept can be adapted much easier to fit another vehicle platform.

Longitudinal control for trajectory tracking itself can be handled in different ways, too. One common approach is to separate trajectory tracking and vehicle dynamics control by applying an inner and outer control loop in a cascade structure, see e.g. \cite{zhu2017}. The inner control loop thereby can be seen as an individual system with input-output behaviour from the outer loop's perspective. The task of the outer loop in our case is to track the position reference trajectory for the vehicle. Due to the dynamic constraints of the vehicle, model predictive control (MPC) is particularly suitable because it can explicitly deal with constraints imposed on the controlled system \cite{rawlings2017}.

MPC also allows to not only consider current trajectory points but also future reference inputs. This can be exploited to achieve a smooth drive while also keeping the longitudinal position error low. In the approach presented in this work, we use multiple models for different dynamical states of the vehicle and thus obtain a hybrid system \cite{MPC_HySys}. The plant-model mismatch shall be minimized this way. Due to the high complexity of the physical model of a powertrain, the models are obtained through input/output (I/O) data identification, providing the behaviour of the inner loop resp. the vehicle dynamics. Despite the usage of several models, a plant-model mismatch will remain and has to be considered in the process of controller design. Since reliability and robustness are crucial for autonomous driving, a stable and fail-safe controller needs to be designed. Besides positional accuracy and before-mentioned goals, the controller should also deliver a smooth and comfortable ride. Previous works dealt with the robust lateral position tracking of a vehicle \cite{RMPC_lat_pos} or robust adaptive cruise control, i.e. the velocity is tracked \cite{RhMPC_ACC}. In the latter work, multiple prediction models were used in the design of the controller. These models were obtained by linearising the nonlinear system equations.
In \cite{RMPC_PWA_BD} the authors also deal with robust MPC and multiple models and apply a similar approach as presented in this paper. However, in \cite{RMPC_PWA_BD} the goal is state regulation (as opposed to output trajectory tracking) and the considered systems are delay-free. 

Therefore, the main contribution of this paper is twofold: First, we use system identification to generate multiple linear time-invariant systems with input delay at different regions of the state space. Second, for this hybrid system a robust model predictive controller is designed in order to track the longitudinal position trajectory of the vehicle. In this context it is shown how the constraints of the optimization problems have to be chosen in order to guarantee robust constraint satisfaction and robust recursive feasibility, especially at switching instances. To the best of the authors' knowledge the combination of those two theoretical aspects applied to the tracking of the longitudinal position of a vehicle instead of the velocity has not been tackled in research yet.

The remainder of this paper is structured as follows. First, a brief introduction to the vehicle setup is given. Then, it is shown how the models for our MPC approach are obtained and transformed to be used within robust MPC. The optimization problem (OP) and its constraints are derived. Next, simulation results are given and discussed. Lastly a conclusion and brief outlook are given. 

\begin{figure*}
	\centering
	\includegraphics[width=1.0\linewidth]{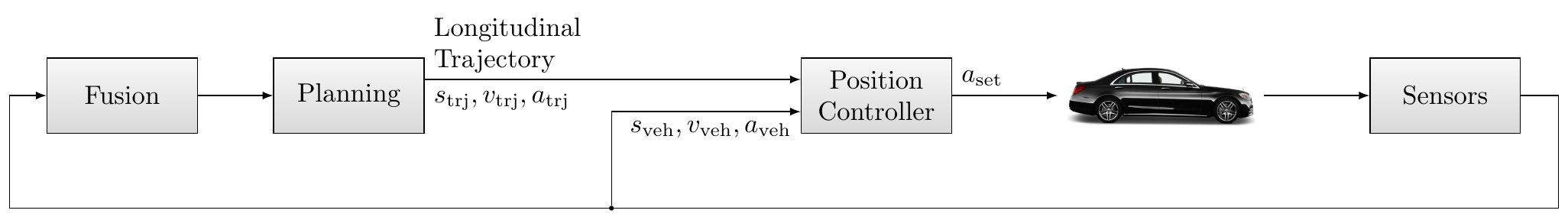}
	\caption{Considered vehicle setup.}
	\label{fig:setup}
\end{figure*}

\textbf{Notation:}
The vertical concatenation of the two column vectors $x,y$ is denoted with $(\vec{x}^T, \vec{y}^T, \vec{z}^T)^T \triangleq \left[\vec{x}, \vec{y}, \vec{z}\right]$.
A matrix $M$ is called Schur stable if $|\eig(M)| < 1$.
The intersection of sets $\mathbb{M}_{i,j}$, where $i=1,\ldots,i_{\max}$ and $j=1,\ldots,j_{\max}$, is denoted with $\bigcap_{i,j} \mathbb{M}_{i,j}$.
The dimension of a vector $\vec{x}\in\R^n$ is provided through $\dim(\vec{x})=n$.
Given two sets $\mathbb{U}$ and $\mathbb{V}$, such that $\mathbb{U} \subset \R^{n}$ and $\mathbb{V} \subset \R^{n},$ the Minkowski sum is defined by $\mathbb{U} \oplus \mathbb{V} \triangleq \{\vec{u}+\vec{v}\!: \vec{u} \in \mathbb{U}, \vec{v} \in \mathbb{V}\},$ the Pontryagin set difference is: $\mathbb{U} \ominus \mathbb{V} \triangleq \{\vec{u}\!: \vec{u} \oplus \mathbb{V} \subseteq \mathbb{U}\}$.

\section{Vehicle setup}
The considered vehicle is a modified sedan which consists of several modules. Relevant modules for longitudinal control are shown in Fig. \ref{fig:setup}. 
A planning module provides a longitudinal reference trajectory containing distance to be travelled $s_{\text{trj}}$, speed reference $v_{\text{trj}}$, acceleration reference $a_{\text{trj}}$ and the corresponding time stamps. This information is extracted from a global position trajectory and serves as an input to the top level position controller. Position trajectory control is chosen over path tracking combined with speed tracking to avoid high frequency replanning. The longitudinal position controller is subject of this work and generates an acceleration demand $a_{\text{set}}$ for a low level torque controller (not depicted) within the vehicle. The vehicle states $s_{\text{veh}}$, $v_{\text{veh}}$, $a_{\text{veh}}$ are measured and fed back to the controllers and to a fusion module which in turn provides combined sensor information to the planning module. Like this, the control loops are closed and the vehicle can act autonomously.
\section{Modelling}
\subsection{System identification}
Models used in our MPC are based on system identification using I/O data of the real vehicle. The data was collected during several test drives and contains a total of more than 900 hours of driving data on proving grounds, including rural roads. Therefore common dynamic requirements for autonomous driving are represented within the dataset. We identify the I/O behaviour of the vehicle acceleration $a_{\text{set}} \rightarrow a_{\text{veh}}$ using an \textbf{A}uto \textbf{R}egressive \textbf{M}oving \textbf{A}verage model with e\textbf{X}ogenous input (ARMAX model). Written as a difference equation, the general model structure is the following:
\begin{equation}
\begin{aligned}
y(t) + &a_1y(t-1)+\ldots+a_{n_a}y(t-n_a) = \\ 
&b_1 u(t-n_k)+\ldots+b_{n_b} u(t-n_k-n_b) +\\ 
&c_1 e(t-1) + \ldots + c_{n_c} e(t-n_c)+e(t)
\end{aligned}
\label{eq:ARMAXlong}
\end{equation}
with $ n_a $, $ n_b $, $ n_c $ representing the number of poles, zeros and error term coefficients, while $ n_k $ covers the input delay. The term $e(t)$ represents a white noise disturbance.
The parameters describing our system behaviour were determined using a brute force approach based on resulting model properties. Those properties are model design parameters and in our case cover the normalized root mean squared error (NRMSE) between model output and real data and whether the resulting model is minimum phase or not. Multiple models describing the dynamic behaviour are obtained by dividing the data dependent on the vehicle's velocity and acceleration. It is assumed that a model identified for a specific region of the vehicle's dynamic is more accurate than an averaged model of the whole dynamics applied to the same specific region. Therefore, with multiple models, a better modelling accuracy can be achieved.

\begin{figure}
	\centering
	\includegraphics[width=0.9\linewidth]{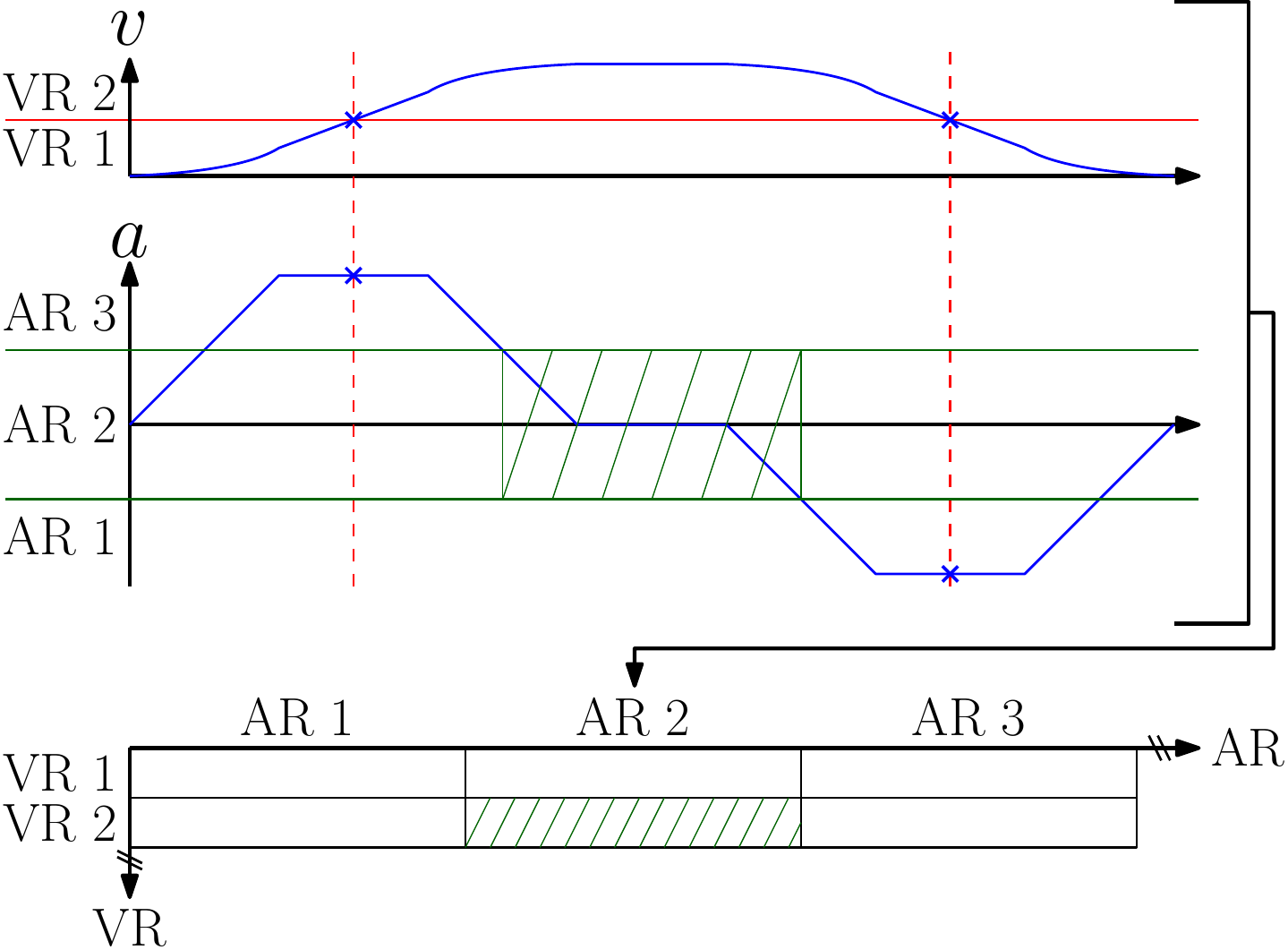}
	\caption{Partitioning the measurement data into velocity and acceleration regions.}
	\label{fig:idReg}	
\end{figure}

This results in data sets for specific acceleration regions ($AR$) within specific velocity regions ($VR$) and allows the system identification of those subsets. An exemplary illustration is given in Fig. \ref{fig:idReg}. 
In this case, data within the marked area belongs to $VR_{i=2}$ and $AR_{j=2}$. Whenever a subset contains not enough data for a meaningful identification, a backup system identified from all available data is used. Although MPC is applicable with an ARMAX model already, in our case it is necessary to transform the systems into state space representation due to our switching strategy when dealing with several models. Representation (\ref{eq:ARMAXlong}) can be transformed into a state space representation with input delay
\begin{equation}
\begin{aligned}
\vec{x}_{\text{id}}(t+1)&=A\vec{x}_{\text{id}}(t)+Bu(t-n_k),
\\
y(t)&=C_{\text{id}} \vec{x}_{\text{id}}(t).
\end{aligned}
\label{eq:IDSS}
\end{equation}

\subsection{System description without input delay}
The identified single input single output (SISO) system \eqref{eq:IDSS} can be transformed such that the first state corresponds to the acceleration of the vehicle. By observing that the transformed output matrix has to fulfil
\begin{equation}
\begin{aligned}
y(t) = a(t) &= C_{\text{id}}\vec{x}_{\text{id}}(t) 
\\
&= \underbrace{C_{\text{id}}T^{-1}}_{\textstyle =:\bar{\bar{C}}_{a}}
\underbrace{T \vec{x}_{\text{id}}(t)}_{\textstyle =:\bar{\bar{\vec{x}}}_{a}(t)}
= \(1, 0, \ldots, 0\) \bar{\bar{\vec{x}}}_{a}(t)
\end{aligned}
\nonumber
\end{equation}
one can deduce that $C_{\text{id}}=\bar{\bar{C}}_{a}T=\(1, 0, \ldots, 0\)T=\(t_{1,1}, t_{1,2}, \ldots\)$. This requirement is met with the regular transformation matrix
\begin{equation}
T =
\def\arraystretch{1.5}
\left( 
\begin{array}{c}
C_{\text{id}}\\ \hdashline[3pt/1.5pt]
\(\vec{0}_{n_{\text{id}}-1\times 1},\, I_{n_{\text{id}}-1}\)
\end{array}
\right).
\nonumber
\end{equation}
By integrating the first state of $\bar{\bar{\vec{x}}}_a=[a, \bar{\bar{x}}_2, \ldots]$ twice, the system can be augmented with the states for position and velocity, $\bar{\bar{\vec{x}}}_{s,v,a}=[s,v,\bar{\bar{\vec{x}}}_{a}]$. The output matrix is such that the first three states are measured, i.e. $\vec{y}=[s,v,a]$.

In order to consider the remaining plant-model mismatch a disturbance term $\bar{\bar{\vec{w}}}_{s,v,a} = [0, 0, w_a, 0, \ldots, 0] \in \bar{\bar{\mathbb{W}}}_{s,v,a} = \{\vec{0}\} \times \mathbb{W}_a \times \{\vec{0}\}$, where $\mathbb{W}_a= [w_{a,\min}; w_{a,\max}]$, is added to the state space model. This yields the system
\begin{equation}
\bar{\bar{\Sigma}}_{s,v,a}
\left\{
\begin{aligned}
&\begin{alignedat}{3}
\bar{\bar{\vec{x}}}_{s,v,a}(t+1) 
&= \bar{\bar{A}}_{s,v,a} \bar{\bar{\vec{x}}}_{s,v,a}(t) 
&&+ \bar{\bar{B}}_{s,v,a} \bar{u}(t-d) 
\\
&&&+ \bar{\bar{\vec{w}}}_{s,v,a}(t)
\\
\vec{y}_{s,v,a}(t) &= \bar{\bar{C}}_{s,v,a} \bar{\bar{\vec{x}}}_{s,v,a}(t)
\end{alignedat}
\\[0.1cm]
& \text{where } \ 
\begin{alignedat}[t]{3}
\bar{\bar{\vec{x}}}_{s,v,a} & \in \bar{\bar{\mathbb{X}}}_{s,v,a}, 
\quad 
&\bar{\bar{\vec{w}}}_{s,v,a} &\in \bar{\bar{\mathbb{W}}}_{s,v,a},
\\
\bar{u} & \in \bar{\mathbb{U}}, 
&\Delta \bar{u} &\in \Delta \bar{\mathbb{U}},
\end{alignedat}
\end{aligned}
\right.
\nonumber
\end{equation}
where $d=n_k$.
At this point one needs to introduce the state and input constraints. Neither the position nor the non-physical states should be constrained. Defining $\tilde{n}:=\dim(\bar{\bar{\vec{x}}}_{s,v,a})-3$, the state constraints are $\bar{\bar{\mathbb{X}}}_{s,v,a}:= \R \times \mathbb{X}_{v,a} \times \R^{\tilde{n}}$. The constraint sets $\mathbb{X}_{v,a},\bar{\mathbb{U}}$ and $\Delta\bar{\mathbb{U}}$ have to be chosen such that they represent the physical limitations of the vehicle.

Following the procedure depicted in \cite{santos2012}, the input delay is shifted to the output altering the state constraint set and the disturbance set. The result is referred to as system $\bar{\Sigma}_{OD,s,v,a}$, which has formally no input delay. In opposite to the procedure where the input delay is removed by augmenting the state vector with memory states the before-mentioned procedure does not increase the system order. This fact is important for the determination of robust positively invariant (RPI) sets which become computationally infeasible for an increasing dimension of the system\todo{ref?}.
\\
Omitting the subscript $(\cdot)_{OD,s,v,a}$ one can introduce the state and input
\begin{equation}
\vec{x}(t) := 
\begin{pmatrix}
\bar{\vec{x}}(t)\\ 
\bar{u}(t-1)
\end{pmatrix}
,\qquad 
u(t):= \Delta\bar{u}(t)=\bar{u}(t)-\bar{u}(t-1)
\nonumber
\end{equation}
which allows to reformulate system $\bar{\Sigma}$ with $u$ as an input:
\begin{equation}
\hspace{-0.7cm}
\Sigma
\left\{
\begin{aligned}
&\begin{alignedat}{3}
\def\arraystretch{1.5}
\underbrace{
	\left( 
	\begin{array}{c}
	s \\ \hdashline[3pt/1.5pt]
	\breve{x}
	\end{array}
	\right)}_{=\vec{x}}(t+1) 
&= 
\def\arraystretch{1.5}
\underbrace{
	\left( 
	\begin{array}{c;{3pt/1.5pt}c}
	A_{11} & A_{22}\\ \hdashline[3pt/1.5pt]
	\vec{0} & \breve{A}
	\end{array}
	\right)}_{=A}
\vec{x}(t) 
&&+ 
\underbrace{
	\left( 
	\begin{array}{c}
	0 \\ \hdashline[3pt/1.5pt]
	\breve{B}
	\end{array}
	\right)}_{=B}
u(t) 
\\
&&&+ 
\underbrace{
	\left( 
	\begin{array}{c}
	w_s \\ \hdashline[3pt/1.5pt] 
	\breve{\vec{w}}
	\end{array}
	\right)}_{=\vec{w}}(t)
\\
\vec{y}_{s,v,a}(t+d) 
&=
\mathrlap{
\def\arraystretch{1.5}
\left( 
\begin{array}{c}
s \\ \hdashline[3pt/1.5pt]
\vec{y}_{v,a}
\end{array}
\right)(t+d)
=
\underbrace{
	\left( 
	\begin{array}{c;{3pt/1.5pt}c}
	1 & \vec{0} \\ \hdashline[3pt/1.5pt]
	\vec{0} & \breve{C}
	\end{array}
	\right)}_{=C}
\underbrace{
	\left( 
	\begin{array}{c}
	s \\ \hdashline[3pt/1.5pt]
	\breve{\vec{x}}
	\end{array}
	\right)}_{=\vec{x}}(t)
}
\end{alignedat}
\\[0.1cm]
& \text{where } \ 
\begin{alignedat}[t]{1}
\vec{x} &\in \mathbb{X}= \bar{\mathbb{X}} \times \bar{\mathbb{U}} = \R\times\breve{\mathbb{X}}, 
\\
\vec{w} &\in \mathbb{W} = \bar{\mathbb{W}} \times \{0\} = \mathbb{W}_s \times\breve{\mathbb{W}}, 
\\
u &\in \mathbb{U} = \Delta \bar{\mathbb{U}}.
\end{alignedat}
\end{aligned}
\right.
\nonumber
\end{equation}
The above shown partitioning shows that there exists a decoupled subsystem $\breve{\Sigma}$ which plays a major role in the computation of positively invariant (PI) sets for system $\Sigma$. These sets are needed to guarantee robust properties within the MPC scheme.

\section{Derivation of the optimization problem}
The goal is to determine a convex OP
\begin{equation}
	\mathcal{P}
	\left\{
		\begin{aligned}
			\underset{\vec{O}}{\text{min}} \
			J
			&= 
			\frac{1}{2} \ \vec{O}^T H \vec{O}
			\ + \ 
			\vec{f}^T \vec{O}
			\\[0.2cm]
			\text{s.t.} \quad
			A_{\text{ineq}}\vec{O}&\leq \vec{b}_{\text{ineq}}
			\\
			A_{\text{eq}}\vec{O}&= \vec{b}_{\text{eq}}
		\end{aligned}
	\right.
	\nonumber
\end{equation}
whose optimal solution $\vec{O}^*$ can be used to compute an input $\bar{u}$. The MPC scheme should be recursively feasible under certain assumptions and guarantee that the propagation of system $\Sigma$ does not violate any constraints, i.e. robust constraint satisfaction is fulfilled.
\form{We proceed step by step by defining cost functional and constraints.}

\subsection{Derivation of the cost functional}
The computed input should minimize the longitudinal position error of the vehicle by simultaneously achieving a comfortable ride. Therefore the predicted position error $e_s=s_{\text{trj}} - s$, its change $\Delta e_s$ and the acceleration error $e_a=a_{\text{trj}} - a$ should be penalized within the prediction horizon. It can be shown that there exist quantities such that the above mentioned errors can be expressed with
\begin{equation}
	\underbrace{
		\begin{pmatrix}
		\vec{E}_s\\
		\vec{E}_{\Delta s}\\
		\vec{E}_a
		\end{pmatrix}
	}_{\textstyle =: \vec{E}_{\text{all}}}
	=
	\underbrace{
		\begin{pmatrix}
		\vec{R}_s\\
		\vec{R}_{\Delta s}\\
		\vec{R}_a
		\end{pmatrix}
	}_{\textstyle =: \vec{R}_{\text{all}}}
	-
	\underbrace{
		\begin{pmatrix}
		\Lambda_s\\
		\Lambda_{\Delta s}\\
		\Lambda_a
		\end{pmatrix}
	}_{\textstyle =: \Lambda_{\text{all}}}
	\underbrace{
		\begin{pmatrix}
		\vec{x}(t+1)\\
		\vec{x}(t+2)\\
		\vdots\\
		\vec{x}(t+N)
		\end{pmatrix}}_{\textstyle =: \vec{X}}.
		\nonumber
\end{equation}
Neglecting the initial errors at time $t$, the cost functional
\begin{equation}
	\check{J} = \frac{1}{2}\vec{E}_{\text{all}}^T Q \vec{E}_{\text{all}} +  \frac{1}{2} \vec{U} R \vec{U},
	\label{eq:constFun}
\end{equation}
penalizes both the error $\vec{E}_{\text{all}}$ and the input sequence $\vec{U} := [u(t),\ldots,u(t+N-1)]$ with the weighting matrices $Q$ and $R$. 
Since the main focus of this paper does not lie on minimizing the computational effort, for simplicity reasons the control horizon was chosen equal to the prediction horizon.
Since the evolution of the undisturbed system is uniquely determined by an initial state and an input sequence, the cost functional can be expressed with the optimization variable
\begin{equation}
	\vec{O} := 
	\begin{pmatrix}
	\vec{o}_1\\	\vec{o}_2\\ \vec{o}_3\\ \vec{o}_4\\ 
	\end{pmatrix} 
	=
	\begin{pmatrix}
	\vec{U} \\
	\breve{\vec{x}}_{t,\text{meas}} - \breve{\vec{x}}_{t,\text{rob}}\\
	\breve{\vec{x}}_{t,\text{meas}}\\
	\gvec{\theta}\\
	\end{pmatrix}
	\nonumber
\end{equation}
and the current measured (estimated) state and reference. \form{The optimization variable $\vec{o}_4=\gvec{\theta}$ is used to pose the steady states of the system, c.f. second paragraph of subsection \ref{derivation_constraints} or \cite{limon2010}.} Omitting the constant terms in $\check{J}$, the cost functional can be represented by the function $J = \frac{1}{2} \vec{O} H \vec{O} + \vec{f}^T\vec{O}$, where $H>0$, i.e. that $J$ is convex with respect to $\vec{O}$.
\\
\form{It should be noted that $\vec{o}_3$ is constrained by the equality $\vec{o}_3=\breve{\vec{x}}_{t,\text{meas}}$, c.f. end of subsection \ref{derivation_constraints}.}

\subsection{Derivation of the constraints}
\label{derivation_constraints}
%
%
%
\textbf{Robust constraint satisfaction}
\\
The goal is to find a MPC scheme that guarantees the satisfaction of the state and input constraints $\mathbb{X},\mathbb{U}$ for any time $t$ and for any disturbance $\vec{w}\in\mathbb{W}$. The idea of the ``tube-based`` approach is to use an additional error feedback to contain the real system state in a ``tube`` around the nominal state. Looking at the block diagonal structure of system $\Sigma$ and taking into account that the first state $s$ is unconstrained, one can deduce that for any $\vec{w}\in\mathbb{W}$ the constraints of $s$ are satisfied. This fact allows us to consider the breve system $\breve{\Sigma}$ only.

Standard robust tube-based MPC theory in chapter 3 of \cite{rawlings2017} states that if the OP is constraint by 
\begin{equation}
\mathcal{C}_{\text{RCS}}
\left\{
\begin{aligned}
\breve{\vec{x}}_{t,\text{rob}} &\in \breve{\vec{x}}_{t,\text{meas}}\oplus (-\breve{\mathbb{Z}})\\
\vec{x}_t &= [s_{t,\text{meas}}, \breve{\vec{x}}_{t,\text{rob}}]\\
\vec{X} &= \Gamma \vec{U} + \Omega \vec{x}_t\\
\vec{x}(t+i)&\in \mathbb{X}_{\text{tight}}=\R \times \breve{\mathbb{X}} \ominus \breve{\mathbb{Z}} \hspace{0.2cm} &&\mathbb{N}_{[0,N]}\\
u(t+i)&\in \mathbb{U}_{\text{tight}}=\mathbb{U}\ominus \breve{K}\breve{\mathbb{Z}}  &&\mathbb{N}_{[0,N-1]}
\end{aligned}
\right.
\label{cJ_robust_track}
\end{equation}
with 
\begin{equation}
	\Gamma :=
	\begin{pmatrix}
		B 		 & \vec{0} 	  & \cdots    	& \vec{0}\\
		AB 		 & B 		  & \ddots 		& \vdots\\
		\vdots   & \ddots 	  & \ddots    	& \vec{0}\\
		A^{N-1}B & \cdots 	  & AB 			& B\\
	\end{pmatrix}, \quad
	\Omega :=
	\begin{pmatrix}
		A\\
		A^2\\
		\vdots\\
		A^N
	\end{pmatrix}	
	\nonumber
\end{equation}
and the feedback law is given by $u=u^*(1) + \breve{K}(\breve{\vec{x}}_{t,\text{meas}} - \breve{\vec{x}}_{t,\text{rob}}^*)$, the constraints are satisfied robustly. Thereby the asterisk notation denotes the optimal solution. Note that $\breve{\mathbb{Z}}$ is an $\epsilon$-outer approximation of the minimal RPI (mRPI) set for the autonomous, disturbed system $\breve{\vec{x}}^+ = (\breve{A}+\breve{B}\breve{K})\breve{\vec{x}} + \breve{\vec{w}}$, where the system matrix is Schur stable. For instance, the feedback gain matrix can be determined by solving the linear matrix inequality (LMI) presented in \cite{limon2010}. Afterwards $\breve{\mathbb{Z}}$ can be computed using the algorithm of \cite{rakovic2005}. \form{Alternatively one can compute the maximal RPI set according to \cite{kolmanovsky1998}, which then can be shrinked by applying the theory presented in \cite{rakovic2004} resulting in $\breve{\mathbb{Z}}$. The latter was done in this paper.}

\textbf{Robust recursive feasibility}
\\
\cite{limon2010} depicts how the idea of the terminal set has to be extended for the tracking case, i.e. if the reference is not constant. It is shown, that one has to impose the additional constraint
\begin{equation}
\mathcal{C}_{\text{RRF}}
\triangleq
\begin{pmatrix}
\vec{x}(t+N)\\ 
\gvec{\theta}
\end{pmatrix}
\in
\mathbb{O}_{\text{tr}}
\nonumber
\end{equation}
where $\mathbb{O}_{\text{tr}}$ is in general the maximal positively invariant set for tracking $\operatorname{MPI}_{\text{tr}} \(A,B,K_{\text{tr}},M_{\theta},\mathbb{X}_{\text{tight}} \times \mathbb{U}_{\text{tight}} := \mathbb{M}\)$, which is defined by
\begin{alignat*}{2}
\left.
\underbrace{
	\begin{pmatrix}
	\vec{x}\\
	\gvec{\theta}
	\end{pmatrix}
}_{\textstyle =: \vec{x}^a}
\right.
^+
&
\mathrlap{
	=
	\begin{pmatrix}
	A+BK_{\text{tr}} & B L\\
	\vec{0} & I_m
	\end{pmatrix}
	\vec{x}^a
}
\\
\vec{x}^a(t) \overset{\text{!}}&{\in} \mathbb{O}_{\text{tr}} = 
\big\{
&&\vec{x}^a = [\vec{x},\gvec{\theta}]:
\\
&&[&\vec{x},K_{\text{tr}}\vec{x} + L \gvec{\theta}], \, M_{\theta}\gvec{\theta} \ \in \mathbb{M}
\big\}	
\quad \forall t
\end{alignat*}
where $A+BK_{\text{tr}}=:A_{K_{\text{tr}}}$ is Schur stable, all steady states can be posed with $[\vec{x}_s, u_s] = M_{\theta}\gvec{\theta}, \ \gvec{\theta} \subseteq \R^m$ and $L:=\(-K_{\text{tr}}, 1\)M_{\theta}$.
 
The stability of $A_{K_{\text{tr}}}$ results in a bounded set $\mathbb{O}_{\text{tr}}$, i.e. that the position $s$ is bounded. This can result in problems since the position of the vehicle can become arbitrarily large. Therefore one needs to alter the theory presented in \cite{limon2010} for our application.

The only steady state of system $\Sigma$ can be represented using the introduced system $\breve{\Sigma}$, i.e. $M_{\theta}=[M_{\theta,1}, \breve{M}_{\theta}]=[1,\vec{0}]$. Choosing the special structure for the feedback gain matrix $K_{\text{tr}} = (0, \ \breve{K}_{\text{tr}})$ and considering the structure of the system matrices one can derive the augmented system
\begin{equation}
	\begin{aligned}
		\left.\vec{x}^a\right.^+ &=
		\begin{pmatrix}
			A+BK_{\text{tr}} & BL\\
			\vec{0} &1
		\end{pmatrix}
		\vec{x}^a
		=
		\def\arraystretch{1.5}
		\left( 
			\begin{array}{c;{3pt/1.5pt}cc}
			A_{11} & A_{12} & 0 
			\\ \hdashline[3pt/1.5pt]
			\vec{0}   & \breve{A}+\breve{B} \breve{K}_{\text{tr}} & 0
			\\
			0 & \vec{0} & 1
			\end{array}	
		\right)
		\begin{pmatrix}
			s\\ \hdashline[3pt/1.5pt]
			\breve{\vec{x}}\\
			\gvec{\theta}
		\end{pmatrix}.
	\end{aligned}
	\nonumber
\end{equation}
Since $s$ is unconstrained and the lower left block of the matrix is zero it obviously applies that $\mathbb{O}_{\text{tr}}=\R \times \breve{\mathbb{O}}_{\text{tr}}$. Furthermore it can be  seen that for all $\gvec{\theta} \in \R$ the constraints $M_{\theta}\gvec{\theta}=[\gvec{\theta},\vec{0}]\in \R \times \{\vec{0}\} \subseteq \mathbb{M}$ are fulfilled and the searched set can be split up into the three parts $\mathbb{O}_{\text{tr}}=\R\times\breve{\mathbb{O}}\times\R$. The MPI set $\breve{\mathbb{O}}$ of the system $\breve{\vec{x}}^+=(\breve{A}+\breve{B} \breve{K}_{\text{tr}})\breve{\vec{x}}$ subject to the constraints $[\breve{\vec{x}},K_{\text{tr}}\breve{\vec{x}} + \vec{0}] \in \breve{\mathbb{X}}\times\mathbb{U}=:\breve{\mathbb{M}}$ exists and is bounded if $(\breve{A}+\breve{B} \breve{K}_{\text{tr}})$ is Schur stable and $\vec{0} \in \breve{\mathbb{M}}$ according to Theorem 4.1 in \cite{kolmanovsky1998}. The searched MPI set then is $\mathbb{O}_{\text{tr}} = \R \times \breve{\mathbb{O}} \times \R$ and the statement $\breve{\vec{x}}(t+N) \in \breve{\mathbb{O}} \ \Rightarrow \ [\vec{x}(t+N), \gvec{\theta}] \in \mathbb{O}_{\text{tr}}$ is true. 

The interpretation of these results is quite intuitive. Recursive feasibility for system $\breve{\Sigma}$ ensures the stability of the origin $\breve{\vec{x}}=\vec{0}$, i.e. that the vehicle is standing still at a position $s$.

\textbf{Constraints}
\\
It can be shown that there exist quantities such that the above shown constraint sets $\mathcal{C}_{\text{RCS}}$ and $\mathcal{C}_{\text{RRF}}$ of the OP can be posed with
\begin{equation}
\begin{aligned}
A_{\text{ineq}} \, \vec{O} &\leq \vec{b}_{\text{ineq}},
\\
A_{\text{eq}} \, \vec{O} &= \vec{b}_{\text{eq}}.
\end{aligned}
\nonumber
\end{equation}
Both vectors $\vec{b}_{\text{ineq}},\vec{b}_{\text{eq}}$ then are a function of the measured value $\vec{x}_{t,\text{meas}}$. 
\form{In order to circumvent the need of computing the Minkowski sum $\breve{\vec{x}}_{t,\text{meas}}\oplus (-\breve{\mathbb{Z}})$ online at each time instance, the first constraint of $\mathcal{C}_{\text{RCS}}$ can be posed as $\vec{o}_2 \in \breve{\mathbb{Z}}$. Then the initial state $\breve{\vec{x}}_{t,\text{rob}}$ can be obtained by $\vec{o}_3-\vec{o}_2$ since we imposed the equality constraint $\vec{o}_3=\breve{\vec{x}}_{t,\text{meas}}$.}


\section{Robust predictive tracking control for multiple models}
In order to apply the given approach to a multi model system, some additional considerations have to be taken. Those are the adaptions needed for robustness and switching between models. In the following the indices $i,j$ represent the velocity and the acceleration region, respectively. In addition, all quantities related to OP before and after a switch are labelled as $BS$ and $AS$, resp.

\subsection{Switching between models}
The identification approach allows that the resulting models are of different orders and model states are not interpretable in a physical way. In addition, the models $\Sigma_{i,j}$, and hence the OPs $\mathcal{P}_{i,j}$, are in general independent from each other. Therefore, when switching from one model to another, the states of the model to be switched to need to be estimated. This is obtained by estimating the states of all models in parallel using Luenberger observers.

\subsection{Tightened constraint sets}
\label{sec:tight_constr_MM}
\begin{figure}[h]
	\vspace{-0.2cm}
	\centering
	\includegraphics[width=.73\columnwidth]{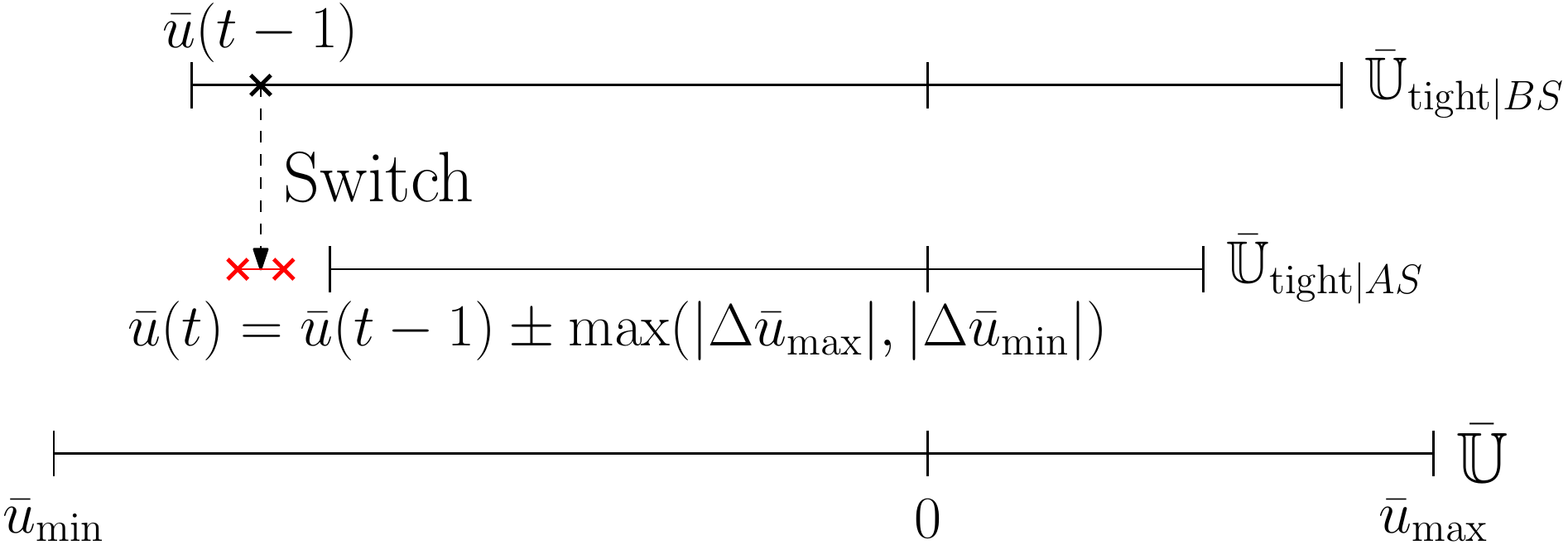}
	\caption{Input and input change constraints cannot be satisfied simultaneously.}
	\label{mm_constr_tight}
\end{figure}

The independences of the models causes the computed quantities $\breve{\mathbb{Z}}_{i,j}$ and $\breve{K}_{i,j}$ to be unequal for different $i,j$.
It follows that in general the tightened state and input constraint sets $\mathbb{X}_{\text{tight}|i,j}$ and $\mathbb{U}_{\text{tight}|i,j}$ are different. This can cause that the constraints are violated when switching to a new model or rather a new OP. No feasible solution can be found. A simple example of how a constraint violation can occur is shown in Fig. \ref{mm_constr_tight}.
Both constraints cannot be satisfied simultaneously, therefore no feasible solution exists.
\\
Since in the set $\mathbb{X}_{i,j}$ neither the position nor the non-physical states are constrained the same applies for
\begin{equation}
\mathbb{X}_{\text{tight}|i,j}= \R \times \mathbb{X}_{\text{tight}|i,j|v,a} \times \R^{n_{i,j}-3} \times \mathbb{X}_{\text{tight}|i,j|\bar{u}}
\nonumber
\end{equation}
A potentially conservative but purposeful solution to this issue is the intersection of all constrained states in $\mathbb{X}_{\text{tight}|i,j}$, i.e. the altered constraint sets are defined by
\begin{equation}
\begin{aligned}
\tilde{\mathbb{X}}_{\text{tight}|i,j} :=& \ \R \times \bigcap\limits_{i,j} \mathbb{X}_{\text{tight}|i,j|v,a} \times \R^{n_{i,j}-3} \times \bigcap\limits_{i,j} \mathbb{X}_{\text{tight}|i,j|\bar{u}}
\\
\tilde{\mathbb{U}}_{\text{tight}|i,j} :=& \ \bigcap\limits_{i,j} 	\mathbb{U}_{\text{tight}|i,j}.
\end{aligned}
\nonumber
\end{equation}

\subsection{Preservation of feasibility at switching points}
Finding a feasible solution for the OP before a switch does not generally imply the same after it. The reason for this are the varying system dynamics, terminal sets for tracking and constraints. In order to deal with this issue, for all possible OPs one has to consider the regions from which the nominal state can be feasibly steered to the terminal set within the prediction horizon. These regions shall be called feasibility sets $\mathcal{F}_{i,j}$. By incorporating the intersection of all feasibility sets in the OPs, which reduces the set of all feasible state sequences because the constraints of the other OPs are implicitly considered, one can guarantee recursive feasibility for the the multi model model predictive tracking control scheme. A sketch of the proof can be found in the \hyperref[appendix]{Appendix}.
\\

\section{Results}
\begin{figure*}[!h]
	\centering
	\begin{subfigure}{.83\textwidth}
		\centering
		\includegraphics[width=\textwidth
		]{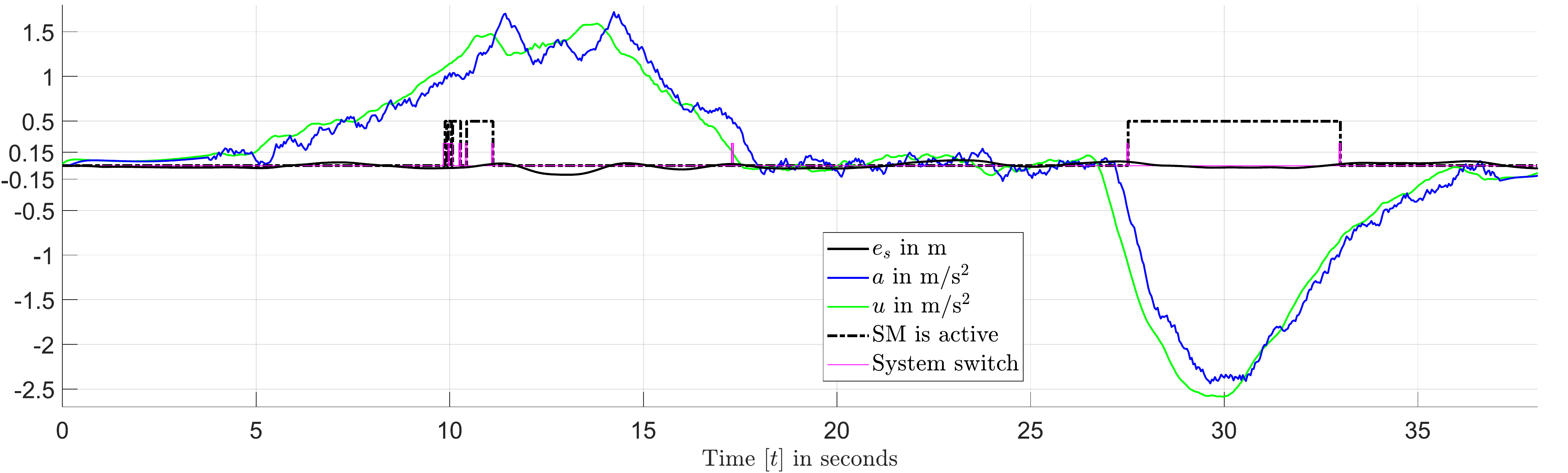}
		\caption{Simulation without additional velocity constraint.}
		\label{fig:resOpt}
	\end{subfigure}
	\begin{subfigure}{.83\textwidth}
		\centering
		\includegraphics[width=\textwidth
		]{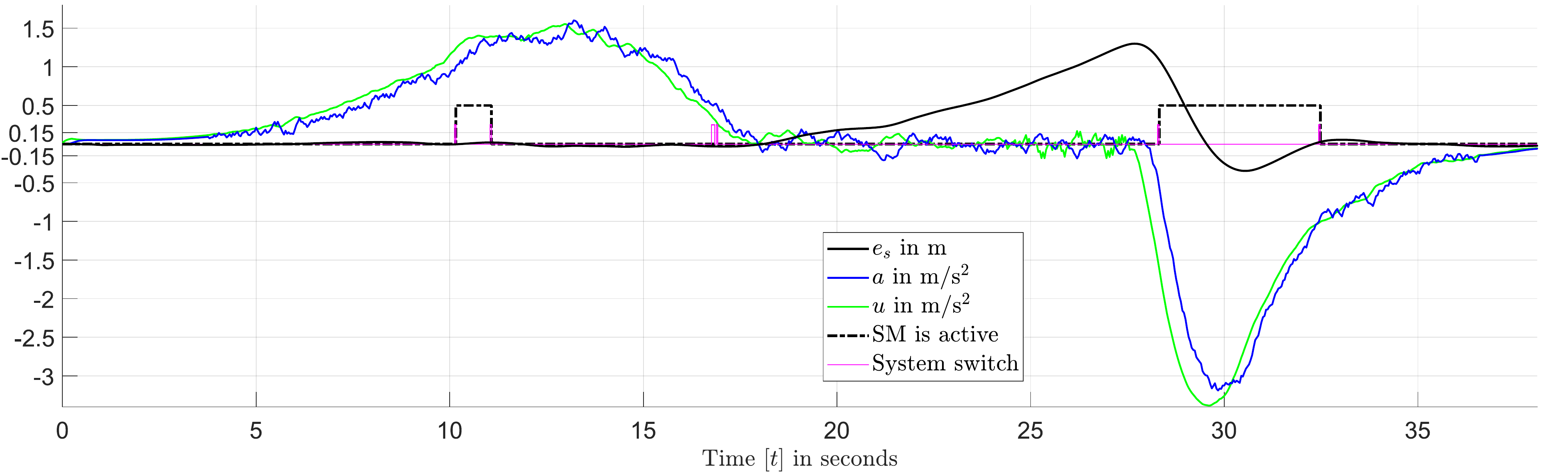}
		\caption{Simulation with additional velocity constraint.}
		\label{fig:resSubOpt}
	\end{subfigure}
	\caption{Simulation results showing controller output $u$ (green), disturbed vehicle acceleration $a$ (blue), position error $e_s$ (black), system switching instants (pink) and whether the backup system is active or not (dashed).}
	\label{fig:results}
\end{figure*}

The proposed controller was tested in simulation using Matlab. By applying the procedure depicted in Fig. \ref{fig:idReg}, five models were identified from real data, including the identified backup system which is denoted as $SM$. The controlled plant is represented by the same models used in the MPC. To test robustness, the acceleration feedback is disturbed through adding random but bounded noise. For comparison reasons, the disturbance was the same for every simulation run. The trajectory to be followed by the MPC was recorded during a test drive. It contains the longitudinal position information fitting to accelerating the vehicle up to a certain speed, keeping that speed constant and then braking to stand still. This behaviour resembles a common situation in city traffic. The weighting matrices of the cost function were determined through a performance criterion and optimization. The prediction horizon $N$ was set to $1$ second or $25$ time steps. In Fig. \ref{fig:results}, two cases are presented. The difference between both is, that for Fig. \ref{fig:resSubOpt} an additional constraint on maximum velocity was added to the OP. In both cases, the disturbance on the vehicle acceleration $a$ can be seen. A switch of models, indicated in pink, does not have a negative impact neither on the control command nor on the acceleration. This is true even for the switch into the backup system, whose model is potentially the least related to the model used before the switch. In total, the result is satisfying, especially when considering that the position error $e_s$ is below 15\unit{cm} in Fig. \ref{fig:resOpt}. The additional velocity constraint, which was set to the maximum velocity within the reference trajectory, changes the OP. As a result, the tightened constraints change as well and for maintaining robustness, the controller limits its control actions in a way that the necessary velocity can not be reached any more. This results in an increasing position error as illustrated in Fig. \ref{fig:resSubOpt}. This is however an issue of position control and requires a trade-off between position accuracy and state constraint satisfaction. It can be seen that the controller recovers fast from this and robustly tracks the trajectory again after a few time instants. By comparing Fig. \ref{fig:resOpt} and \ref{fig:resSubOpt}, it can be seen that different constraints possibly change the control output and therefore the vehicle response, however in both cases, the trajectory remains traceable and additional disturbances are robustly compensated. 

Concerning the real time capability, it is clear that the current implementation is not applicable in real time. Especially when switching to the backup system it can be observed that real time factor spikes. This is a sign that at a switching point, the two considered models differ more than when switching between two systems in neighbouring regions.


%

\section{Conclusion and Outlook}
In this paper we presented an approach on how to apply robust model predictive control on multiple model systems. The method is used to control the longitudinal position of a vehicle. It is shown how different models are obtained from measurement data and necessary system transformations needed for applying robust control theory are described. The approach was successfully tested in simulation, showing satisfying results for position control. Maintaining recursive feasibility despite of switched systems was obtained by introducing feasibility sets and robust constraint satisfaction can be assured through the conservative approach of intersecting tightened constraint sets. \form{Future research therefore should include more rigorous testing is required both through simulations and real-life testing. Eventually, the applicability of the derived additional constraints to guarantee recursive feasibility should be investigated and if there exist less restrictive approaches to do so.}




\IncludeAppendix{
\appendix

\section{Proof}
\label{appendix}
Consider the class of discrete-time switched linear disturbed input delayed systems with bounded input change and constraints on the state and input:
\begin{equation}
	\bar{\bar{\Sigma}}_{\bar{\bar{\sigma}}(\bar{\bar{\vec{x}}}(t))} 
	\left\{
    	\begin{aligned}
      		&\begin{alignedat}{3}
      			\bar{\bar{\vec{x}}}(t+1) &= \bar{\bar{A}}_{\bar{\bar{\sigma}}(\bar{\bar{\vec{x}}}(t))} \bar{\bar{\vec{x}}}(t) &&+ \bar{\bar{B}}_{\bar{\bar{\sigma}}(\bar{\bar{\vec{x}}}(t))} \bar{\bar{\vec{u}}}(t-d)
      							\\ &&&+ \bar{\bar{\vec{w}}}_{\bar{\bar{\sigma}}(\bar{\bar{\vec{x}}}(t))}(t)
	     		\\
	     		\vec{y}(t) &= \bar{\bar{C}}_{\bar{\bar{\sigma}}(\bar{\bar{\vec{x}}}(t))} \bar{\bar{\vec{x}}}(t)
      		\end{alignedat}
      		\\[0.1cm]
      		& \text{where } \ 
      		\begin{alignedat}[t]{3}
	     		\bar{\bar{\vec{x}}} &\in \bar{\bar{\mathbb{X}}}_{\bar{\bar{\sigma}}(\bar{\bar{\vec{x}}}(t))},
		     	\quad 
		     	&\bar{\bar{\vec{w}}}_{\bar{\bar{\sigma}}(\bar{\bar{\vec{x}}}(t))} &\in \bar{\bar{\mathbb{W}}}_{\bar{\bar{\sigma}}(\bar{\bar{\vec{x}}}(t))},
		     	\\
		     	\bar{\bar{\vec{u}}} &\in \bar{\bar{\mathbb{U}}}_{\bar{\bar{\sigma}}(\bar{\bar{\vec{x}}}(t))}, 
	     		&\Delta \bar{\bar{\vec{u}}} &\in \Delta\bar{\bar{\mathbb{U}}}_{\bar{\bar{\sigma}}(\bar{\bar{\vec{x}}}(t))},
      		\end{alignedat}
   		\end{aligned}
	\right.
	\nonumber
\end{equation}
where $\bar{\bar{\vec{x}}}(t) \in \mathbb{R}^{\bar{\bar{n}}_{x}}$ is the state, and $\bar{\bar{\vec{u}}}(t) \in \mathbb{R}^{\bar{\bar{n}}_{u}}$ is the control input; $\bar{\bar{\sigma}}(\bar{\bar{\vec{x}}}(t)):=\{i\in\mathbb{M} \ | \ \bar{\bar{\vec{x}}}(t) \in \mathcal{A}_{i}\}$, where $\mathcal{A}_i$ is the region where system $i$ is ``active``,
is a switching signal that is a piecewise constant function of the current state vector $\vec{x}(t)$, continuous from the right everywhere, and takes values at the sampling times in a finite set $l=\{1, \ldots, M\}=:\mathbb{M}$, where $M>1$ is the number of subsystems.
\\
Now for each subsystem $\bar{\bar{\Sigma}}_{l}, \ l\in\mathbb{M}$, by introducing a new input vector corresponding to the input change and a new state vector containing the current state and the previous input, i.e. $\bar{\vec{u}}(t)=\Delta \bar{\vec{u}}(t)$ and $\bar{\vec{x}}(t):=[\bar{\bar{\vec{x}}}(t), \bar{\bar{\vec{u}}}(t-1)]$, system formulations $\bar{\Sigma}_{l} $
can be derived. In order to formally remove the input delay of the systems $\bar{\Sigma}_{l}$ without introducing memory states, one can follow the explicit dead-time compensation depicted in \cite{santos2012}. The results are system formulations
\begin{equation}
	\Sigma_{l} 
	\left\{
    	\begin{aligned}
      		&\begin{alignedat}{3}
      			\vec{x}(t+1) &= A_{l} \vec{x}(t) + B_{l} \vec{u}(t) + \vec{w}_{l}(t)
	     		\\
	     		\vec{y}(t+d) &= C_{l} \vec{x}(t)
      		\end{alignedat}
      		\\
      		& \text{where } \ 
      		\begin{alignedat}[t]{3}
	     		\vec{x} &\in \mathbb{X}_{l}
	     		= \bar{\mathbb{X}}_{l} \ominus \bar{\mathbb{E}}_{l}
	     		= \(\bar{\bar{\mathbb{X}}}_{l} \times \bar{\bar{\mathbb{U}}}_{l}\) \ominus \bigg(\bigoplus\limits_{k=0}^{d-1} \bar{A}_{l}^k \bar{\mathbb{W}_{l}} \bigg),
		     	\\
		     	\vec{w} &\in \mathbb{W}_{l}= \bar{A}_{l}^d \bar{\mathbb{W}}_{l},
				\quad
		     	\vec{u} \in \mathbb{U}_{l} = \bar{\mathbb{U}}_{l} = \Delta\bar{\bar{\mathbb{U}}}_{l}
      		\end{alignedat}
   		\end{aligned}
	\right.
	\nonumber
\end{equation}
where $\bar{\mathbb{W}}_{l} = \bar{\bar{\mathbb{W}}}_{l} \times \{\vec{0}_{n_u \times 1}\}$ and $\bar{\mathbb{E}}_{l} \subseteq \bar{\mathbb{X}}_{l}$, i.e. $\{\vec{0}\} \subseteq \mathbb{X}_{l}$, by assumption. It can be shown that since all sets for all systems $\bar{\bar{\Sigma}}_{l}$ fulfil all standard properties (containment of the origin and additionally the disturbance set is bounded and convex) by assumption, the same applies for the  sets of $\Sigma_{l}$.
Now by setting the disturbances in all systems $\Sigma_{l}$ to zero, i.e. $\vec{w}_{l}(t)\equiv \vec{0}
$, one can define the nominal systems
\begin{equation}
	\Pi_{l} 
	\left\{
   		\begin{alignedat}{3}
   			\vec{z}(t+1) &= A_{l} \vec{z}(t) + B_{l} \vec{v}(t) 
   			\\
   			\vec{y}(t+d) &= C_{l} \vec{z}(t) 
   			\\[0.1cm]
   			[\vec{z}, \vec{v}] &\in \mathbb{Z}_{l} \times \mathbb{V}_{l} =: \mathbb{L}_{l}
   		\end{alignedat}
	\right.
	\nonumber
\end{equation}
which serve as prediction models for the MPC.
Furthermore it is assumed that for all $l\in\mathbb{M}$ there exists a stabilizing feedback matrix $K_{l}$ and a corresponding minimal RPI set $\mathbb{S}_{l}$ such that the tightened constraint sets $\mathbb{X}_{l}^{\text{t}}=\mathbb{X}_{l}\ominus\mathbb{S}_{l}$, $\mathbb{U}_{l}^{\text{t}}=\mathbb{U}_{l}\ominus K_{L}\mathbb{S}_{l}$ are non-empty. In order to guarantee robust recursive feasibility within each system or rather its corresponding optimization problem (OP) the invariant set for tracking $\mathcal{T}^{\text{aug}}_{l}$, defined in \cite{limon2010}, needs to be incorporated into the OP. Let the current time instance be $t$, then the OP is the following:
\begin{equation}
	\mathfrak{P}_{l}(t)
	\left\{
    	\begin{aligned}
      		&\underset{}{\text{min}} \ J = \vec{E}_{l}^T Q_{l} \vec{E}_{l} + \vec{e}_{l}(0)^T\tilde{Q}_{l}\vec{e}_{l}(0) + \vec{V}^T R_{l} \vec{V} 
      		\\[0.1cm]
      		& \text{s.t. } \ 
      		\begin{alignedat}[t]{1}
      			&\vec{x}(t)-\vec{z}(t|t)			
      			\in \mathbb{S}_{l}, 
				\\
		     	&z(0)=z(t|t),
		     	\\
		     	&\vec{z}(i+1) = A_{l} \vec{z}(i) + B_{l} \vec{v}(i), \ i\in\mathbb{N}_{[0, N_{l}-1]}
		     	\\
		     	&
		     	\begin{pmatrix}
		     		\vec{z}(i) \\ \vec{v}(i)
		     	\end{pmatrix}
		     	\in \tilde{\mathbb{L}}_{l}, \ i\in\mathbb{N}_{[0, N_{l}-1]}
		     	\\
		     	&
		     	\begin{pmatrix}
		     		\vec{z}(N_{l}) \\ \gvec{\theta}	
		     	\end{pmatrix}
		     	\in \mathcal{T}^{\text{aug}}_{l}
      		\end{alignedat}
   		\end{aligned}
	\right.
	\nonumber
\end{equation}
with $\tilde{\mathbb{L}}_{l} = \bigcap_{i\in \mathbb{M}} \mathbb{X}_{i}^{\text{t}}\times\mathbb{U}_{i}^{\text{t}}\subseteq\mathbb{L}_{l}= \mathbb{X}_{l}^{\text{t}}\times\mathbb{U}_{l}^{\text{t}}$, $\vec{E}_{l}= [\vec{e}_{l}(1),\ldots,\linebreak\vec{e}_{l}(N_{l})]$, where $\vec{e}_{l}=\vec{r}-C_{l}\vec{z}$, and $\vec{V}=[\vec{v}_{l}(1),\ldots,\vec{v}_{l}(N_{l})]$.

In contrast to the chosen cost functional in \cite{limon2010}, in our case the optimization variable $\gvec{\theta}$ does not affect the cost. Since $\mathcal{T}^{\text{aug}}_{l}$ is non-empty, there always exist a $\gvec{\theta} \in \operatorname{Proj}_{\gvec{\theta}}(\mathcal{T}^{\text{aug}}_{l})$, hence it can be neglected and the last constraint of $\mathfrak{P}_{l}(t)$ can be altered to $\vec{z}(N_{l})\in \mathcal{T}_{l}:= \operatorname{Proj}_{\vec{z}}(\mathcal{T}^{\text{aug}}_{l})$. If at time $t$ with $\sigma(\vec{x}(t)):=\{i\in\mathbb{M} \ | \ (I_{\bar{\bar{n}}_x}, \vec{0}) \vec{x}(t) \in \mathcal{A}_{i}\}=l$ the optimization problem $\mathfrak{P}_{l}(t)$ has a solution, then robust constraint satisfaction and robust recursive feasibility can be guaranteed for all $p \in  \mathbb{N}_0$ for which $\sigma(\bar{\vec{x}}(t+p))=l$ is true, i.e. until a system switch occurs. It should not go unmentioned that $\forall \ l \in \mathbb{M}$ the OP $\mathfrak{P}_{l}(t)$ can be formulated as a convex constrained OP if the matrices $Q_{l},\tilde{Q}_{l},R_{l}$ are positive definite.

	\addtolength{\textheight}{-4.12cm}    

Definition: 
For an undisturbed system $\Pi_{l}$ the one-step backwards feasibly reachable set of $\mathbb{I}$ is denoted as
\begin{equation}
	\begin{aligned}
			\mathcal{R}\!\operatorname{each}_{l}^{\text{b,f}} \left(\mathbb{I}, \Pi_{l} \right) 
			&\triangleq
			\begin{aligned}[t]
				\big\{
					\vec{z} \in \mathbb{R}^{n} \ \big| \ 
						&\vec{z}_{0} \in \mathbb{I}, \ 
						\vec{z} \in \mathbb{Z}_{l}, \
						\vec{v} \in \mathbb{V}_{l},
						\\
						&\vec{z}_{0}=A_{l}\vec{z} + B_{l} \vec{u}
				\quad
				\big\}
			\end{aligned}
	\end{aligned}
	\nonumber
\end{equation}
The $H$-step backwards feasibly reachable set $\tilde{\mathcal{F}}_{l}^{H}(\mathbb{I},\Pi_{l})$ is defined as
\begin{equation}
	\begin{aligned}
		\tilde{\mathcal{F}}_{l}^{y+1}\(\mathbb{I},\Pi_{l}\) &\triangleq \mathcal{R}\!\operatorname{each}_{l}^{\text{b,f}}
			\(\tilde{\mathcal{F}}_{l}^{y}\(\mathbb{I},\Pi_{l}\), \Pi_{l}\), 
			\quad y \in \mathbb{Z}_{[0, H-1]}
		\\
		\tilde{\mathcal{F}}_{l}^{0}\(\mathbb{I},\Pi_{l}\) &= \mathbb{I}
	\end{aligned}
	\nonumber
\end{equation}

\newtheorem{theorem}{Theorem}
\begin{theorem}
	Under the system dynamics and constraints of $\Pi_{l}$ the set $\mathcal{T}_l$ is feasibly reachable from all $\vec{z}(t) \in \mathcal{F}_{l}$ in $N_{l}$ steps, i.e. $\forall \vec{z}(t)\in  \mathcal{F}_{l} \ \exists \vec{v}(t+i)\in \mathbb{V}_l, \ i\in \mathbb{N}_{[0,N_{l}-1]}$ such that $\vec{z}(t+i)\in \mathbb{Z}_l, \ i\in \mathbb{N}_{[0,N_{l}]}$ and $z(t+N_{l})\in\mathcal{T}_l$,  if	$\mathcal{F}_{l} = \tilde{\mathcal{F}}_{l}^{N_{l}}(\mathcal{T}_{l},\Pi_{l})$.
\end{theorem}

\noindent\textbf{Proof:}
\begin{adjustwidth}{0.2cm}{}
	The existence of a feasible state and input sequence directly follows from the definition of the $H$-step backwards feasibly reachable set. The feasibility of the last state $\vec{z}(t+N_{l})\in\mathcal{T}_{l}\subseteq\mathbb{Z}_{l}$ results from the definition of the (maximal) positively invariant set $\mathcal{T}_{l}$.
\end{adjustwidth}

Let $\mathcal{F}_{\text{all}}:= \bigcap_{i\in \mathbb{M}} \mathcal{F}_{i}$. Consider the extended optimization problem $\mathfrak{P}_{l}^{\text{ext}}(t)$, which is equal to $\mathfrak{P}_{l}(t)$ with the additional constraints
\begin{equation}
	\mathcal{C}_{\text{MMRRF}}\!: \
	\vec{z}(i) \in \mathcal{F}_{\text{all}}, \ i \in \mathbb{N}_{[0,N]},
	\quad
	\vec{z}(1) \in \mathcal{F}_{\text{all}} \ominus \mathbb{S}_{l}.
	\nonumber
\end{equation}

\begin{theorem}
	Assume that at the current time ${t}$ the state $\vec{x}(t)=\bar{\vec{x}}(t+d|t)$ and therefore the current OP $\mathfrak{P}_{l}^{\text{ext}}(t)$, where $l=\sigma(\vec{x}(t))$, is known. Further assume that $\mathfrak{P}_{l}^{\text{ext}}(t)$ is feasible.
	Then there exists a feasible solution for $\mathfrak{P}_{\sigma(\vec{x}(t+1))}^{\text{ext}}(t+1)$.
\end{theorem}


\noindent\textbf{Proof:}
	\begin{enumerate}
		\setlength\itemsep{0em}	
		\item Assume that $\sigma(\vec{x}(t+1))=l$. Then no switch occurred and the \textit{standard} OP $\mathfrak{P}_{l}(t+1)$ is feasible (following the \textit{standard} argumentation with candidate solution and terminal set) because $\mathfrak{P}_{l}^{\text{ext}}(t)$ was feasible by assumption. The feasibility of $\mathfrak{P}_{l}^{\text{ext}}(t+1)$ follows from the fact that optimal state sequence of $\mathfrak{P}_{l}(t+1)$ is contained in $\mathcal{F}_{l}\subseteq\mathcal{F}_{\text{all}}$ .
		\item Assume that $\sigma(\vec{x}(t+1))=k\neq l$. The tube-based approach guarantees $\vec{x}(t+1) \in \vec{z}(t+1|t) \oplus \mathbb{S}_{l}$. By choosing $\vec{z}(0)=\vec{z}(t+1|t+1)=\vec{x}(t+1)$ the constraint $\vec{x}(t+1)-\vec{z}(t+1|t+1)=0\in \mathbb{S}_{k}$ is fulfilled $\forall k \in \mathbb{M}$.
		Furthermore $\vec{z}(0)=\vec{z}(t+1|t+1)=\vec{x}(t+1)\in \vec{z}(t+1|t) \oplus \mathbb{S}_{l} = \vec{z}(1) \oplus \mathbb{S}_{l} \subseteq \mathcal{F}_{\text{all}} \ominus \mathbb{S}_{l} \oplus \mathbb{S}_{l} \subseteq \mathcal{F}_{\text{all}} \subseteq \mathcal{F}_{k}$, i.e. there exists a feasible input sequence such that the initial nominal state $\vec{z}(0)$ can be feasibly steered to the terminal set $\mathcal{T}_{k}$. All constraints are fulfilled, hence $\mathfrak{P}_{\sigma(\vec{x}(t+1))}^{\text{ext}}(t+1)$ has a feasible solution. 
	\end{enumerate}
}

\bibliographystyle{ieeetran}
\bibliography{bib}

\end{document}